\documentclass[useAMS,usenatbib]{mn2e}
\usepackage[dvips]{graphicx}
\usepackage{amsmath}
\usepackage{epsfig}
\def\apj{ApJ }

\def\aj{AJ }
\def\apjl{ApJL }
\def\aj{AJ }

\title[An automated archival VLA transients survey.]{An automated archival VLA transients survey.}
\author[M.E.Bell et al.]{M.E. Bell$^{1}$\thanks{E-mail: meb1w07@soton.ac.uk (MEB)}, R.P. Fender$^{1}$, J. Swinbank$^{2}$, J.C.A. Miller-Jones$^{3,4}$, C.J. Law$^{5}$,
\newauthor
B. Scheers$^{2,8}$, H. Spreeuw$^{2,6}$, M.W. Wise$^{6,2}$, B.W. Stappers$^{7}$, R.A.M.J. Wijers$^{2}$,
\newauthor J. Hessels$^{6}$ and J. Masters$^{3}$.\\
$^{1}$Department of Physics and Astronomy, Southampton University, UK.\\
$^{2}$Astronomical Institute ``Anton Pannekoek'', University of Amsterdam, Science Park 904, 1098 XH, Amsterdam, The Netherlands.\\
$^{3}$NRAO Headquarters, 520 Edgemont Road, Charlottesville, VA 22903, USA. \\
$^{4}$ICRAR -- Curtin University of Technology, GPO Box U1987, Perth, WA 6845, Australia.\\
$^{5}$Radio Astronomy Lab, University of California, Berkeley, CA, 94720, USA.\\
$^{6}$ASTRON, PO Box 2, 7990 AA Dwingeloo, The Netherlands.\\
$^{7}$Jodrell Bank Centre for Astrophysics, University of Manchester, Oxford Road, Manchester, UK.\\
$^{8}$Centrum Wiskunde and Informatica (CWI), PO Box 94079, 1090 GB, Amsterdam, The Netherlands.\\}

\begin{document}

\date{Accepted 2011 March 02. Received 2010 October 22}

\pagerange{\pageref{firstpage}--\pageref{lastpage}} \pubyear{2011}

\maketitle

\label{firstpage}

\begin{abstract}
In this paper we present the results of a survey for radio transients using data obtained from the Very Large Array archive. We have reduced, using a pipeline procedure, 5037 observations of the most common pointings -- i.e. the calibrator fields. These fields typically contain a relatively bright point source and are used to calibrate `target' observations: they are therefore rarely imaged themselves. The observations used span a time range $\sim$ 1984 -- 2008 and consist of eight different pointings, three different frequencies (8.4, 4.8 and 1.4 GHz) and have a total observing time of 435 hours. We have searched for transient and variable radio sources within these observations using components from the prototype LOFAR transient detection system. In this paper we present the methodology for reducing large volumes of Very Large Array data; and we also present a brief overview of the prototype LOFAR transient detection algorithms. No radio transients were detected in this survey, therefore we place an upper limit on the snapshot rate of GHz frequency transients $>$8.0 mJy to $\rho \leq$ 0.032 deg$^{-2}$ that have typical timescales 4.3 to 45.3 days. We compare and contrast our upper limit with the snapshot rates -- derived from either detections or non-detections of transient and variable radio sources  -- reported in the literature. When compared with the current Log N -- Log S distribution formed from previous surveys, we show that our upper limit is consistent with the observed population. Current and future radio transient surveys will hopefully further constrain these statistics, and potentially discover dominant transient source populations. In this paper we also briefly explore the current transient commissioning observations with LOFAR, and the impact they will make on the field.
\end{abstract}


\section{Introduction}

The scope and depth of transient radio science is vast: by utilising the time domain we can gain unique insight into such objects as neutron stars and white dwarfs in binary systems, relativistic accretion and consequent jet launch around black holes, distant gamma-ray burst afterglows, supernovae and active galactic nuclei (AGN), to name a few. The distances to these objects, as well as the timescale for transient behaviour, varies dramatically.
For example, giant kilo-Jansky micro-second radio pulses have been observed from the relatively nearby Crab Pulsar (e.g. see \citealt{GP}). In contrast, month timescale (and longer) variations are commonly observed in the radio emission produced by powerful jets driven by accretion onto super-massive black holes in distant AGN (\citealt{devries}, \citealt{Me}, \citealt{Sadie}).
Through studying the transient and variable nature of these exotic and energetic objects, we obtain an unprecedented laboratory to probe extreme physics. 

Despite the scientific potential, the transient and time variable sky is a relatively unexplored region of parameter space. 
Historically radio transient detections have been sparse due to an inefficient survey figure of merit to adequately sample a large amount of sky to sufficient sensitivity and time resolution \citep{Cordes_2004, Hessels_LOFAR}. Some detections of transients have therefore been made serendipitously (for examples see \citealt{Davies}, \citealt{Zhao}, \citealt{Flare_oort}, \citealt{Bower_Orion} and \citealt{Lenc_Trans}). This limitation will soon be relieved by the next generation of wide field telescopes and their respective dedicated transient surveys. 

A variety of new wide-field facilities will soon be available to sample the transient sky.
In the optical band the Palomar Transient Factory (PTF; \citealt{PTF}) and Panoramic Survey Telescope and Rapid Response System (Pan-STARRS; \citealt{PanSTARRS}) will survey the sky for transients. 
In the radio band the Allen Telescope Array (ATA; \citealt{ATA}), the Murchison Wide Field Array (MWA; \citealt{MWA}) and the Low Frequency Array (LOFAR; \citealt{Fender_2008}) will soon begin or have already commenced operations. Other wide field radio pathfinders such as the Karoo Array Telescope (MeerKAT; \citealt{MeerKAT}) and the Australian Square-Kilometer-Array Pathfinder (ASKAP; \citealt{ASKAP}) are also being developed on the road to the Square Kilometer Array (SKA). Transient studies are a key science goal for all of these facilities. 

\begin{table*}
\centering
\caption{Number of images reduced and searched with respect to observing frequency. Note, 4.8 and 8.4 GHz were the primary frequencies of interest, 1.4 GHz images were only produced for one field. The calibrators are referred to by their J2000 epoch name. $<\!\delta T_{next}\!>$ is the average change in time between sequential observations (including observation on the same day -- see section 2.1). $<\!\tau\!>$ is the average integration time spent on the calibrator.}
\begin{tabular}{|c|c|c|c|c|c|c|c|c|c|c|}
\hline
\hline  Field & R.A. & Dec. & $\ell$ & \textit{b} & $\#$ Obs. & $\#$ Obs. & $\#$ Obs. & $<\!\delta T_{next}\!>$ & $<\!\tau\!>$ & \\ 
  & (J2000) & (J2000) & & & (8.4 GHz) & (4.8 GHz) & (1.4 GHz)  & (days) & (mins)& \\ 
\hline
1800+784 & 18$^{h}$00$^{m}$45$^{s}$.7 & +78$^{\circ}$28$^{m}$04$^{s}$.0 & 110.0 & +29.1 & 992 & 908  & 151 & 4.3 & 5.6 &  \\ 
0508+845 & 05$^{h}$08$^{m}$42$^{s}$.4 & +84$^{\circ}$32$^{m}$04$^{s}$.5 & 128.4 & +24.7 & 205 & 413 & - & 13.6 & 5.1 & \\ 
1927+739 & 19$^{h}$27$^{m}$48$^{s}$.5 & +73$^{\circ}$58$^{m}$01$^{s}$.6 & 105.6 & +23.5 & 171 & 19 & - & 45.3 & 9.0 & \\ 
1549+506 & 15$^{h}$49$^{m}$17$^{s}$.5 & +50$^{\circ}$38$^{m}$05$^{s}$.8 & 80.2  & +49.1 & 480 & 558 & - & 8.3 & 5.0 & \\ 
0555+398 & 05$^{h}$55$^{m}$30$^{s}$.8 & +39$^{\circ}$48$^{m}$49$^{s}$.2 & 171.6 & +7.2  & 123 & 190 & - & 27.3 & 3.5 & \\ 
2355+498 & 23$^{h}$55$^{m}$09$^{s}$.5 & +49$^{\circ}$50$^{m}$08$^{s}$.3 & 113.7 & -12.0 & 183 & 99 & - & 29.3 & 7.6/15.3$^{a}$ & \\
3C48     & 01$^{h}$37$^{m}$41$^{s}$.3 & +33$^{\circ}$09$^{m}$35$^{s}$.1 & 25.1  & +33.4 &  -  & 545&   & 16.2 & 4.6 & \\
\hline  Total & & & & & 2154 & 2732 & 151 & 1.9/3.0$^{b}$ & 5.2 & \\ 
\hline 
\end{tabular} 
\begin{flushleft}
$^{a}$ \textit{7.6 mins at 8.4 GHz and 15.3 mins at 4.8 GHz.} \\
$^{b}$ \textit{1.9 days including observations on the same day / 3.0 days disregarding them.} 
\end{flushleft}
\end{table*}

A common method to detect radio transients is through multi-wavelength triggered observations from, for example, all sky monitors on X-ray observatories. These have produced radio counterparts to gamma-ray burst (GRB) afterglows and black hole X-ray binary outbursts (for examples see \citealt{Frail_GRB}, \citealt{gaensler_magnetar}, \citealt{eck_SN}). This method relies on having a detectable high frequency counterpart, which may be absent (or difficult to detect) for sources such as X-ray dim isolated neutron stars (XDINs; \citealt{Ofek}) and orphan gamma-ray burst afterglows \citep{Frail_GRB}, demonstrating the need for dedicated radio transient programs. 

Despite the historical challenges, dedicated or commensal transients surveys have produced a number of interesting results.
The Galactic centre (GC) has been the area for some intense observing campaigns; these studies have so far detected a number of radio transients, the most recent being GCRT J1742-3001 \citep{hyman_recent_galactic_2009} and GCRT J1746-2757 \citep{Hyman_2002_GC} -- also see \cite{Bower_GC_2005}.
In the high time resolution domain, \cite{RRAT_1} discovered short duration transient radio bursts from neutron stars, i.e. RRATs (Rapidly Rotating Transients) -- as well as many detections of new pulsars (for example, from the Parkes multi-beam survey). 

Nine bursts -- named the WJN transients -- in excess of 1 Jy have been discovered using drift scan observations with the Waseda Nasu Pulsar Observatory at 1.4 GHz (summarised in \citealt{Matsumura_2009}, but also see \citealt{Kuniyoshi_2007}, \citealt{Niinuma_2007}, \citealt{kida}, \citealt{Niinuma_2009} for further details). These are some of the brightest transients reported in the literature and so far remain unexplained. 
Recently \cite{Croft} published results from the ATA Twenty Centimetre Survey (ATATS): no transients were detected and an upper limit on the snapshot rate of events was given. Subsequently the Pi GHz Sky Survey (PiGSS) surveyed the sky with the ATA at 3.1 GHz, providing the deepest static source catalogue to date above 1.4 GHz \citep{Bower2010}. No transients were reported in this survey and an upper limit on snapshot rate was placed. 

Searching for highly variable \textit{known} radio sources can also be a useful diagnostic in examining the dynamic radio sky. 
For example, \cite{Carilli} found a number of highly variable ($\Delta S \geq\pm 50$\%) radio sources in a small number of repeated observations of the Lockman Hole at 1.4 GHz. \cite{Frail_CAT} also found four highly variable radio transient sources from follow-up observations of GRBs at 5 and 8.5 GHz: the rates of these events are consistent with those reported in \cite{Carilli}. Reporting 39 variable radio sources, \cite{Becker} recently characterised the surface density of variables in the direction of the Galactic plane at 4.8 GHz. Most of the variable sources detected had no known multi-wavelength counterparts. This is an important study as the rates of transient and variable sources may differ in the direction of Galactic plane when compared with an extra-Galactic pointing. In particular, large numbers of flare stars are known to produce bright coherent bursts \citep{Flare_stars}, and could dominate detections at low frequencies \citep{Bast_Flare}. A deeper discussion on the difference between transient and variable processes will follow later.  

Radio telescope archives potentially contain many hours of data which currently remains unsearched for radio transients.  
An archival study comparing the NVSS (NRAO VLA Sky Survey; \citealt{Condon_NVSS}) and FIRST (Faint Images of the Radio Sky at Twenty-cm;
\citealt{Becker_FIRST}; \citealt{White_FIRST}) catalogues was conducted by Levinson et al. (2002), with a follow-up study by Gal-Yam et al.(2006): a number of radio transient sources were identified.
\cite{Bower_2007} analysed 944 epochs of archival VLA data at 4.8 and 8.4 GHz spanning a period of 22 years. In this survey ten radio transients were reported, with the host galaxies possibly identified for four out of the ten sources, and the hosts and progenitors of the other six unknown.  
\cite{Keith} recently published results from a search for transient and variable sources in the Molonglo Observatory Synthesis Telescope (MOST) archive at 843 MHz.
15 transient and 53 highly variable sources were detected over a 22 year period. \cite{Keith} use these detections to place limits on the rates of transient and variable sources. 
\cite{BowerCAL} have published further archival work examining observations of the calibrator 3C286 at 1.4 GHz. A total of 1852 epochs are examined spanning a time range 23 years: no radio transients were reported.

\cite{Bower_2007} with its high yield of transients provided the motivation for this work.
The aim of this paper is to push archival radio transient studies further, within the framework of testing and refining the transient detection algorithms that will operate on the LOFAR radio telescope \citep{Fender_2008}.
We present the findings of an 8.4, 4.8 and 1.4 GHz study of the repeatedly observed flux and phase calibrator fields found in the VLA archive. 
We explore from some of the publications discussed above the reported snapshot rates of either detections of radio transients, or upper limits based on non-detections: we place our own upper limit on these values and discuss the implications.

\begin{figure}
\label{Hist}
\includegraphics[scale=0.88]{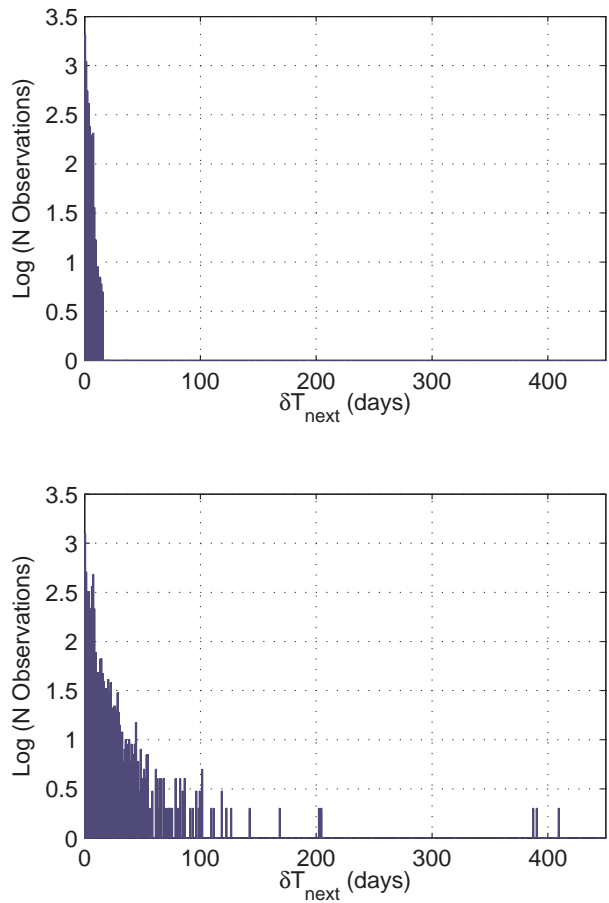}
\caption{Top Panel: Histogram showing the time difference between observations -- for \textit{all} pointings and frequencies within this survey. Bottom Panel: A (summed) histogram of the time differences between observations \textit{per pointing} including all frequencies. For a description of the average cadence per pointing see table 1.}
\end{figure}

\section{VLA Data}
\subsection{VLA calibrator fields}
We have searched part of the VLA archive for transients. 
In order to optimise our chances of success we searched for the most repeatedly observed fields in the VLA archive.
The flux and phase calibrator fields were chosen as the backbone of this new transient study. These calibrators are observed routinely and are a standard and necessary calibration technique in radio interferometry. The calibrator fields usually contain a relatively bright compact object, typically a quasar. The calibrators were selected to fulfil one or more of the following criteria: (A) they were observed frequently; (B) they should be unresolved on the longest A-configuration VLA baseline; (C) they had a relatively large integration time per observation.
The chosen VLA fields are summarised in Table 1. Note that we initially focused our efforts on the flux calibrators, specifically 3C48; however, with the nature of the bright source in the field and the typically short integration time ($\sim$ 1 minute) the images often suffered from artifacts. Therefore we quickly switched our attention to the phase calibrators -- which proved, due to longer integration times ($\sim 5$ mins) to have better image fidelity. 

A total of 5037 flux and phase calibrator images have been reduced and searched, with a total observing time of 435 hours. The average integration time spent on all sources was $<\!\tau\!>$ = 5.2 mins. A full statistical description of the measured noise per image is given in section 5. 

The mean separation between observations regardless of pointing and frequency was $<\!\delta T_{next1}\!>$ = 1.9 days. 
When calculating this average we have included observations that occurred on the same day. Note that it is quite common for an observation of the same source to be performed at a number of different frequencies (i.e. sequentially within the same observation). This obviously produces a bias that reduces the average time between observations. When producing the images for this survey we only logged the date of the observations, not the exact start and stop time. Extracting the start and stop time was not easily executed within the imaging pipeline framework. We therefore add in an arbitrary time delay of 4 hours in calculating the averages for observations that have the same date. Ignoring observations on the same date yields an average $<\!\delta T_{next2}\!>$ = 3.0 days (regardless of pointing and frequency). 

See the top panel of Figure \ref{Hist} for a histogram of the time differences between sequential observations for \textit{all} pointings and frequencies. The bottom panel of Figure \ref{Hist} shows a (summed) histogram of the time differences between observations \textit{per pointing}.
Table 1 summarises the average time difference between observations and the average integration time per pointing. The lowest average time difference between observations achieved for one single pointing was 4.3 days (1803+784); the highest was 45.3 days (1927+739). 
As we have sampled a number of different fields at different cadences we state the range 4.3 to 45.3 days as the timescale of transient behaviour that we would be sensitive to. In quoting these numbers we do make the assumption that each image has an equal chance of making a detection -- i.e that there is an isotropic distribution of radio transients and there is no frequency dependence (between 1.4 and 8.4 GHz) for detection. 

\subsection{VLA data analysis}
The archival data have been reduced using the ParselTongue (see \citealt{Kettenis}) Python interface to the Astronomical Image Processing System (AIPS; \citealt{AIPS}). A scripted procedure was written to perform the following tasks: 
\begin{itemize}
\item The data were loaded into AIPS.
\item The antenna table was searched for antennas that were `out' or designated `EVLA' (only relevant 2006 onwards) during the observations. These antennas were then flagged from being used in subsequent calibration and imaging tasks. 
\item The maximum baseline length in the observation was ascertained; this was then utilised to optimise cell and image sizes respectively in the subsequent imaging steps. 
\item The automated flagging procedures were applied to flag unwanted and erroneous visibilities. 
\item The observation log was searched for \textit{any} known flux or phase calibrator\footnote{Found at http://www.aoc.nrao.edu/$\sim$gtaylor/csource.html} used at the VLA within an entire observation.
\item For \textit{any} identified calibrator, standard calibration was performed.
\item The calibrator was subsequently imaged and deconvolved, followed by three iterations of phase self-calibration only, and then one iteration of amplitude and phase self calibration. The time interval for phase calibration was set appropriately with respect to the integration time on source.
\item The calibrator was boxed off, modelled and removed from the image using UV component subtraction.
\item Finally the subtracted image was lightly CLEANed with 150 iterations \citep{CLEAN}.
\end{itemize}

The script was designed to be run on large volumes of data without interruption. Python exception handling was used to catch potential errors and remove bad data from further processing. An example image of 3C48 produced by the pipeline before and after source subtraction is shown in Figure 2.  Both images show contours and Grey scale to give the reader an intuitive feel of the image quality. The source to the North of 3C48 is persistent with a flux $\sim$ 30 mJy and will be discussed further in section 4. 

\section{Transient Search} 
The images produced by the imaging pipeline were then processed through the prototype LOFAR transient detection algorithms \citep{Swinbank}.
By prototype, we refer to a well tested subset of algorithms (taken from LOFAR) that were put together in a `pipeline' to perform the source extraction and databasing for this transient search only. It should be noted that a broader set of algorithms with greater functionality will define the final LOFAR transient detection system. 

For each image a background RMS map was calculated over the entire image. For any island of pixels above 8$\sigma$ i.e. eight times the noise measured from the RMS map, source extraction was performed by fitting elliptical Gaussians. For further details on the LOFAR source extraction algorithms see Spreeuw (2010).
In the case of LOFAR images, a subsection of the entire image will be used to calculate a `local' RMS map. Considering the increased field of view of LOFAR, coupled with greater source counts, it is not sufficient to define a global RMS map over the entire image. 

\begin{figure*}
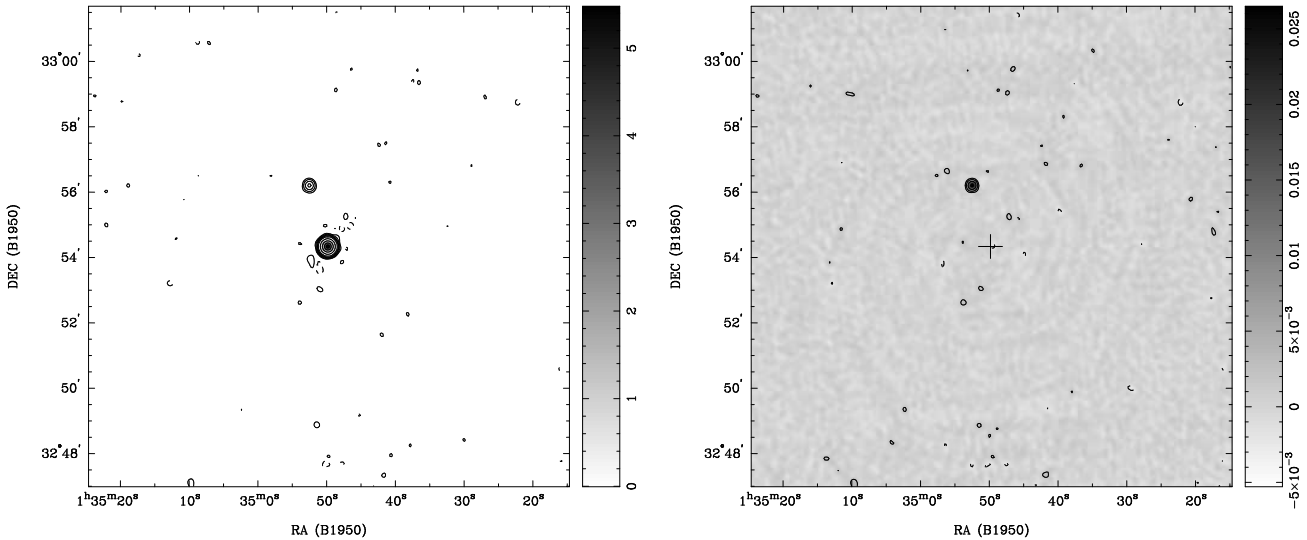

\label{3C48}
\centering
\begin{tabular}{cc}
\epsfig{file=Figure2_1.ps,width=0.4\linewidth,angle=270,clip=} &
\epsfig{file=Figure2_2.ps,width=0.4\linewidth,angle=270,clip=} \\
\end{tabular}
\caption{An example image of 3C48 at 4.8 GHz produced by the pipeline (observation date 1984-10-05). The left panel shows a CLEANed image before source subtraction. The right shows the same epoch after source subtraction, a cross denotes the original position of 3C48. The wedge on the right of each image shows the intensity of the Grey scale in Janskys. The integration time was $\sim$4 mins, yielding an RMS of 0.6 mJy.
Contour levels are -3, 3, 8, 20, 40, 100, 500, 2000, 5000, 8000 $\times$ RMS for both images.}
\end{figure*}

\begin{figure}
\includegraphics[scale=0.8]{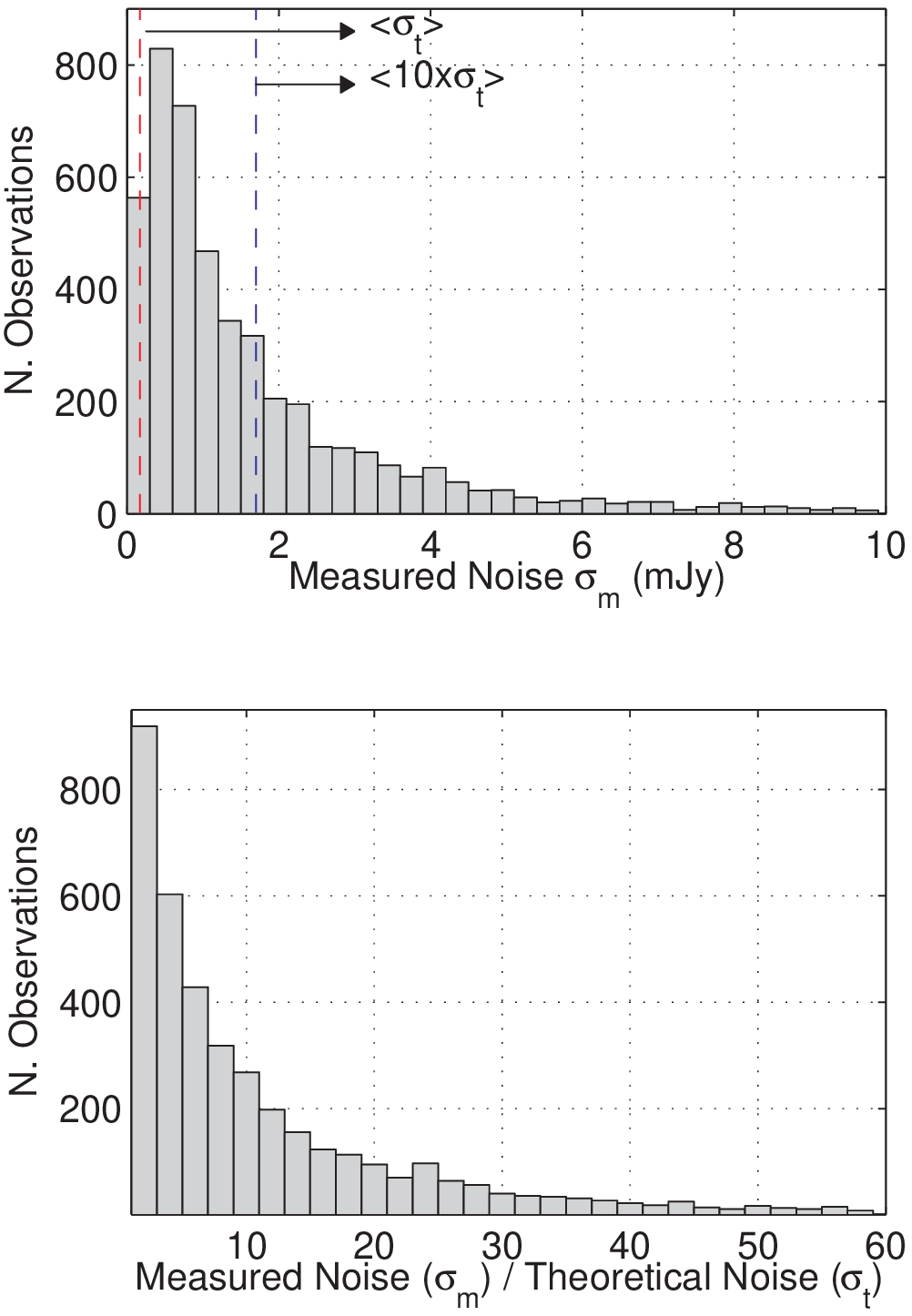}
\label{Sens_plot}
\caption{Top panel: Histogram showing the measured noise -- calculated in the same region -- for all images. $<\sigma_{t}>$ gives the theoretical noise calculated from the average integration time for all observations; $<10\times\sigma_{t}>$  gives this limit multiplied by 10. Bottom panel: Histogram showing the measured noise divided by the theoretical per observation -- which accounts for different integration times.}
\end{figure}

A conservative threshold of $8\sigma$ was chosen after initial tests and pipeline refinements indicated that a badly calibrated and reduced image -- considering the typical strength of the calibrator flux -- could produce a large number of artifacts and thus false source detections.
All images that were processed by the imaging pipeline were processed through the prototype LOFAR transient detection algorithms -- errors included. 
The rationale for this was to explore the effects of badly calibrated images with respect to source extraction and transient detection. LOFAR will incorporate a false detection rate (FDR) algorithm in the source extraction system, where the global detection threshold for source extraction is set to minimise the number of false positives and is governed by the individual image statistics \citep{FDR}.

After source extraction a MonetDB (Boncz 2002) database was then populated with the measured properties and associated data of the extracted sources.
The source properties include: position and associated errors, all Stokes parameters of peak and integrated flux including the Gaussian fitting parameters.
The associated data included, for example, time of observation, operating frequency and beam properties. 
The database was searched for either unique sources that had no previously known counterparts (in previous images), or a known source whose flux had varied by a significant amount. We adopt the same metric used in \cite{Carilli} to define significant variability as $\Delta S \geq \pm$50\%. 

For further details on the transient database algorithms, including a mathematical description of the variability measure used in this survey see Scheers (2010). The database will also cross-reference the transient parameters with those in the WENSS (Westerbork Northern Sky Survey; \citealt{WENSS}), NVSS and VLSS (VLA Low-frequency Sky Survey; \citealt{VLSS}) catalogues to search for counterparts; for this survey we typically relied on NVSS. 

For the majority of fields in this survey, after the calibrator source was subtracted, the fields were left almost devoid of sources: thus the transient search was relatively trivial. A few $\sim$mJy radio sources were present in some of the fields and only some of the time, due to changing sensitivity. Therefore, we concentrated our efforts on locating unique sources, rather than characterising the variability of known sources. 
Once the final list of candidate transient sources had been produced, light curves were automatically generated and the images of interest were checked for calibration errors and image fidelity. 

\section{Results}
A total of 5037 images at various pointings and frequencies have been searched at a detection level of $8\sigma$.
Nine candidate transients that were detected in images with adequate image fidelity were scrutinised further. Four of these candidates were found to lie consistently on the dirty beam. After re-reduction they were shown to be calibration errors which had been cleaned to a point source. 

One candidate was detected 76 times at 4.8 and 8.4 GHz and was found to be significantly variable. 
After careful consideration and review of the literature it was concluded that this source was created by a bug in the VLA recording system, whereby the pointing of telescopes was changed without updating the header information; this error had affected a previously reported transient VLA J172059.9+385226.6 (see \citealt{Ofek} for further details).
This error was \textit{not} detected in any of the other calibrator images taken around the same time.

Three of the candidates were found to be associated with known weaker radio sources. These sources were detected in the deepest observations ($\sim$ 30 mins) of the respective fields. As observations at this depth were very sparse, these radio sources were considered -- by the database algorithms -- as transient. These candidates were quickly removed when cross referenced with the radio catalogues.
The last transient candidate although surviving some of the re-reduction tests was discarded after it was re-calibrated and the significance level dropped below adequate levels.

The 3C48 field did contain a persistent source at $\alpha=01^{h}37^{m}44^{s}.2$ and $\delta=+33^{\circ} 11^{m}26^{s}.5$ (J2000) with $S_{v} \sim$ 30 mJy. This source could \textit{not} be identified in the NVSS catalogue, due to insufficient resolution to separate it from 3C48. The FIRST survey did not cover the position needed to catalogue this source. The source is however previously identified in high dynamic range studies of 3C48 (see \citealt{Briggs}).  
This source was searched for significant variability but none was found. 
 
\section{Surface Density Upper Limit}
As we have detected no radio transients with this survey, we use the area surveyed per observation, coupled with the typical sensitivity to constrain the snapshot rate of transient events.
To calculate the 2$\sigma$ upper limit of the snapshot rate of transients from our survey we assume a Poisson distribution; for zero detections (n=0) we use:
\begin{equation}
P(n) = e^{-\rho N}
\end{equation}
where $\rho$ is the snapshot rate of transients; and the $2\sigma$ confidence interval is defined as $P(n) = 0.05$ at the 95\% confidence level. $N$ is the sum of the number of images multiplied by the field of view ($\Omega$) at that given frequency i.e.
\begin{equation}
N = (\Omega_{8.4}\times N_{8.4})+(\Omega_{4.8}\times N_{4.8})+(\Omega_{1.4}\times N_{1.4})
\end{equation}
Note, we only consider a search area within the half-power radius per image. Evaluating equations 1 and 2 yields a snapshot rate of $\rho \leq$ 0.032 deg$^{-2}$. 

To evaluate the flux density limit that we are sensitive to when searching for transients, we must statistically consider the noise measured in all the images. Figure 3 presents this information in two different ways. Firstly, on the top panel we show a histogram of the measured noise $\sigma_{m}$ in all the images; included on the plot is an indicator of the theoretical noise in a 5 minute observation (0.16 mJy at 4.8 GHz), and also ten times this value.
Secondly -- in the bottom panel -- we show a histogram of the measured noise divided by the theoretical noise $\sigma_{t}$ (i.e. $\sigma_{m}/\sigma_{t}$) for all images: thereby taking into consideration that not all observations were $\sim$ 5 minutes in duration. 

\begin{figure*}
\includegraphics[scale=0.63]{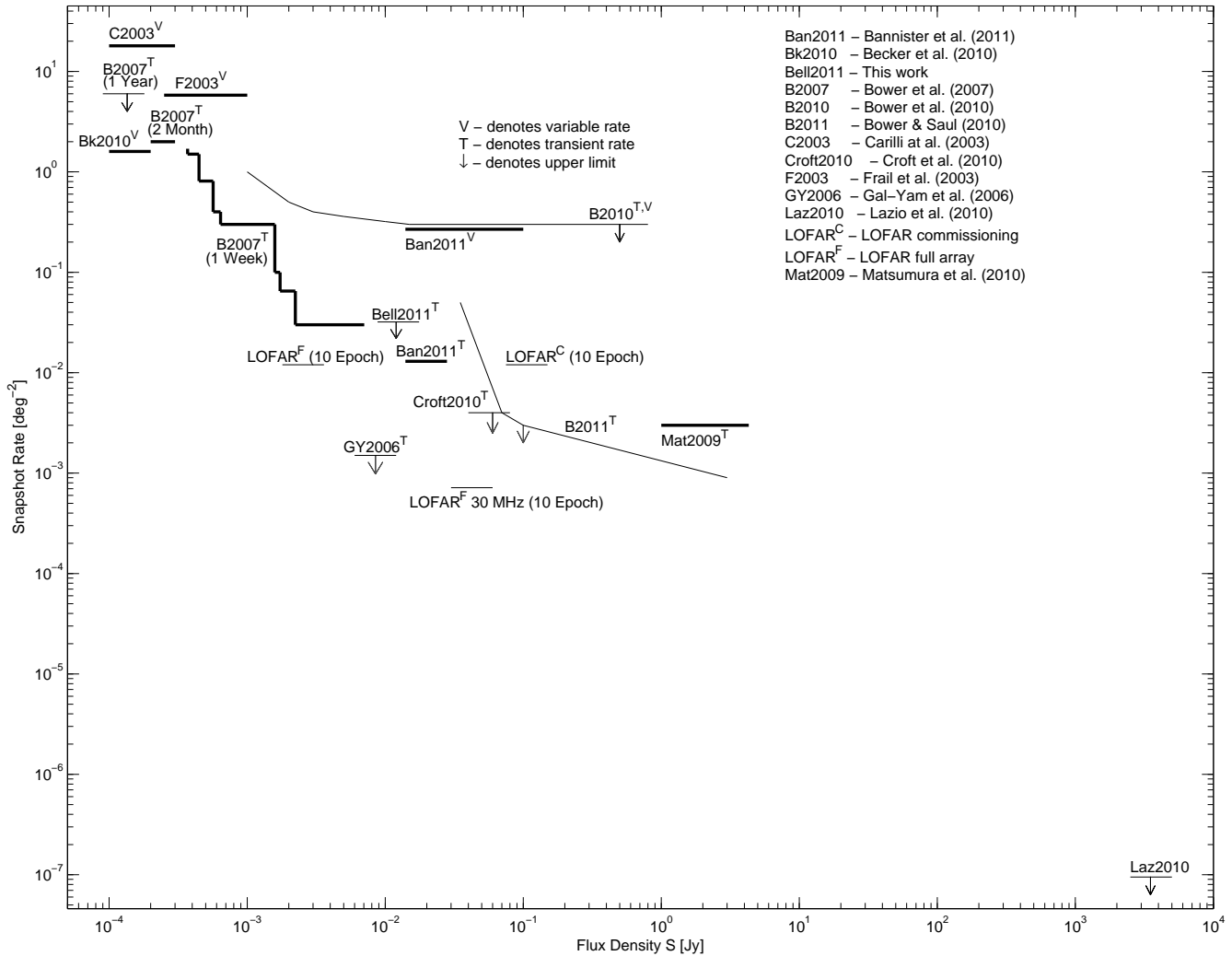}
\label{Flux_vs_ArealDens}
\caption{Snapshot rate (deg$^{-2}$) against flux density (Jy) of detections of transients (labelled `T'), detections of variable sources (labelled `V') and upper limits based on non-detections (labelled with downward arrows). The thick black line denotes detections; the thin line denotes upper limits. The Bower et al. (2007) 1 week, year and two month limits are indicated as B2007$^{T}$ with the appropriate time-scale. 
Ban2011$^{V}$ and Ban2011$^{T}$ indicates the separate rates derived for variables and transients reported in \citet{Keith}.
Bell2011$^{T}$ indicates the 2$\sigma$ upper limits derived from this study. `LOFAR$^{F}$' indicates the theoretical constraint that LOFAR could provide with zero detections from 10 epochs of 12 hour observations, each of 25 deg$^{2}$ fields (using 18 core and 18 remote stations at 150 MHz); `LOFAR$^{C}$' indicates the current commissioning capabilities at 150 MHz. `LOFAR$^{F}$ 30 MHz' shows the rate calculated for a 30 MHz field of view (assuming the final theoretical noise is reached with 18 core and 18 remote stations).  
We note that this plot does not contain any information on characteristic time duration and recurrence of transient behaviour as both are currently poorly constrained.}
\end{figure*}

\begin{table*}
\centering
\caption{Summary of snapshot rates reported in the literature. The results are separated out according to upper limits based on non-detections (top); transient detections (middle); and detections of highly variable radio sources (bottom). The flux min column designates the detection threshold of the observations reported in the literature or the minimum flux of detections (indicated as such);  the maximum flux is only indicated for transient detections. The \citet{Bower_2007} results have been stated three times depending on the characteristic time scale sampled. We do not give the number of epochs for the WJN transients as it is not stated clearly in the literature. \citet{Bower2010} and \citet{BowerCAL} state two different rates depending on flux density, we quote these separately as (A) and (B).}
\begin{tabular}{|c|c|c|c|c|c|c|c|c|c|}
\hline
\hline  Survey/Paper & Flux Min & Flux Max  & $\rho$ & $t_{char}$ & $\nu$ & Epochs&\\ 
&  ($\mu$Jy) & ($\mu$Jy) & (deg$^{-2}$) &  & (GHz) & (N)&\\ 
\hline 
This work & $>$8000 (8$\sigma)$ & - & $<$0.032 & 4.3 - 45.3 days & 8.4, 4.8 and 1.4 & 5037 & \\
FIRST-NVSS/\cite{Gal-Yam} & $>$6000$^{a}$ & - & $<$1.5$\times10^{-3}$ & - & 1.4 & 2$^{b}$ &\\
ATATS/\cite{Croft} & $>$40000 & - & $<$0.004 & 81 days - $\sim$ 15 years & 1.4 & 12$^{b}$ &\\
\cite{Bower_2007} & $>$90 & - & $<$6 & 1 year & 4.8 and 8.4 & 17 &\\
PiGSS-I/\cite{Bower2010}(A) & $>$1000 & - & $<$1 & 1 month & 3.1 & 75 & \\
PiGSS-I/\cite{Bower2010}(B) & $>$10000 & - & $<$0.3 & 1 month & 3.1 & 75 & \\
\cite{BowerCAL}(A) & $>$70000 & - & $<$3$\times10^{-3}$ & 1 day & 1.4 & 1852 & \\
\cite{BowerCAL}(B) & $>$3$\times10^{6}$ & - & $>$9$\times10^{-4}$ & 1 day & 1.4 & 1852 & \\
\cite{Lazio2011} & $>$2.5$\times10^{9}$(5$\sigma$) & - & $<$9.5$\times10^{-8}$ & 5 mins & 0.0738& $\sim$1272 & \\
\hline
\cite{Keith} & 14000(5$\sigma$) & 6.5$\times10^{6}$ & $1.3\times10^{-2}$ & days - years &  0.843 & 3011$^{b}$ &\\
\cite{Bower_2007} & 370 & 7042 & 1.5$\pm$0.4 & 20 mins - 7 days & 4.8 and 8.4 & 944 &\\
\cite{Bower_2007} & 200 & 697 & 2 & 2 months & 4.8 and 8.4 & 96 &\\
WJN/\cite{Matsumura_2009} & 1$\times10^{6}$ & 4.3$\times10^{6}$ & 3$\times10^{-3}$ & $\sim$ 1 day & 1.4 & - & \\
\hline
\cite{Keith} & $>$14000 & - & 0.268 & days - years &  0.843 & 3011$^{b}$ &\\
\cite{Carilli} & $>$100 & - & $<$18 & 19 days and 17 months & 1.4 & 5 &\\
\cite{Becker} & $>$100 & - & 1.6 & $\sim$ 15 years & 4.8 & 3$^{b}$ &\\
\cite{Frail_CAT} & $>$250 & - & 5.8 & $\sim$ 1 day & 5 and 8.5 & - &\\
\hline 
\hline
\label{density_table}
\end{tabular} 

\begin{flushleft}
$^{a}$ \textit{Different noise values were found in each survey map thus global threshold taken above 6 mJy}\\
$^{b}$ \textit{Combined mosaic}\\
\end{flushleft}
\end{table*}

It can be seen that the bulk of the images achieved an image noise less than $10\sigma_{t}$. Possible deviations away from the theoretical noise could be attributed to, for example, unremoved RFI, bad calibration solutions, effects of a bright source in the field and, in general, settings and assumptions within the imaging pipeline that do not lend themselves to a given observation.
Taking the median value of all the measured noises we find $\sigma_{median}$ = 1 mJy (or 6.25$<\!\sigma_{t}\!>$). Using 8$\sigma_{median}$ as the global detection threshold for the entire survey, we find that we would be sensitive to transients $>$8 mJy, with typical timescales 4.3 to 45.3 days (see section 2.1).

In calculating this upper limit we have included \textit{all} images reduced by the pipeline; however, not all images were reduced successfully. 
Although some images contained artifacts, they rarely contaminated the entire image, thus some area could still be searched effectively. If a unique transient point source was detected in an image with errors, the image was re-reduced by hand and checked for reproducibility (see section 4).

In Figure 4 we compare the limit imposed on the snapshot rate of sources from this study, with those found in the literature -- this figure is derived from Fig. 9 of \citet{Bower_2007} and Fig. 20 of \citet{Croft}. We do not include a typical timescale of the transient duration in the snapshot rate calculation as it is not well constrained. This information is summarised and referenced in Table \ref{density_table} and described further in the following paragraphs. 

The \cite{Bower_2007} survey reported the snapshot rate of transients to be $\rho$=1.5$\pm$0.4 deg$^{-2}$ (labelled `B2007$^{T}$ 1 Week' in Figure 4) from eight detections, with characteristic timescale 20 minutes $< t_{char} <$ 7 days, above a flux density 370 $\mu$Jy (with typical image noise $\sim$ 50 $\mu$Jy at the pointing centre). Two transients were detected in the 2 month averaged images above a flux density of 200 $\mu$Jy, giving a 2$\sigma$ limit on the snapshot rate $\rho\sim$2 deg$^{-2}$ (labelled `B2007$^{T}$ 2 Month' in Figure 4). No transients were detected in the year long averages above 90 $\mu$Jy: limiting the 2$\sigma$ snapshot rate to $\rho<$6 deg$^{-2}$ (labelled `B2007$^{T}$ 1 Year' in Figure 4). 

The PiGSS-I survey using the ATA at 3.1 GHz sets an upper limit on the snapshot rate of transients to $\rho<1$ deg$^{-2}$ at 1 mJy, and $\rho<0.3$ deg$^{-2}$ at 10 mJy (labelled `B2010$^{T,V}$' in Figure 4), with characteristic timescale one month \citep{Bower2010}.
A recent study by \cite{BowerCAL} - using archived VLA observations of the flux calibrator 3C286 at 1.4 GHz - set an upper limit of $\rho<3\times10^{-3}$ deg$^{-2}$ at 70 mJy, and $\rho<9\times10^{-4}$ deg$^{-2}$ at 3 Jy (labelled `B2011$^{T}$'). 
The work of \cite{BowerCAL} is very comparable to the work presented in this paper, however, in this study by using predominantly the phase calibrator fields, we slightly push the mean sensitivity down. 

The \cite{Carilli} study set an upper limit on the rate of highly variable radio sources $\geq$ 100 $\mu$Jy to $<$ 18 deg$^{-2}$ with characteristic timescales of 19 days and 17 months (labelled `C2003$^{V}$' in Figure 4). Note that these were detections of variable radio sources. \cite{Frail_CAT} derived a comparable quantity to \cite{Carilli} of $\rho\sim$ 5.8 deg$^{-2}$ with four highly variable sources with characteristic timescale $\sim$ 1 day, above a flux density 250 $\mu$Jy (labelled `F2003$^{V}$' in Figure 4).
Similar to these surveys but in the direction of the Galactic plane \cite{Becker} found 39 variable radio sources between a flux density range 1 to 100 mJy, varying on timescales of years: they derived $\rho\sim$ 1.6 Galactic sources deg$^{-2}$ (labelled `Bk2010$^{V}$' in Figure 4). 

In the context of this survey, the difference between a variable and a transient -- from a purely observational sense -- is a matter of detectability: transients sit, most of the time, below the detection capabilities of the instrument; while variables sit above or close to it. 
However, the underlying astronomical processes associated with transient and variable sources may differ, and should require a different treatment when considering the rates of events.
For example, we might expect the rates of variable sources to differ from `one off' explosive transients such as GRB afterglows -- which have a finite lifetime and will be undetectable beforehand. 
For future surveys, a spectrum of transient and variable behaviour will be observed depending on the cadence and sensitivity. The boundaries between the definitions will become more blurred as the cadence and sensitivity is increased.  

The WJN transients summarised in \cite{Matsumura_2009} range in flux density from 1 to 4.3 Jy with characterstic timescale $\sim$ 1 day, yielding a snapshot rate $\rho\sim$3$\times10^{-3}$ deg$^{-2}$ (labelled `Mat2009$^{T}$' in Figure 4). 
In comparison the \cite{Croft} survey set a 2$\sigma$ upper limit on the snapshot rate of events $>$ 40 mJy to be $\rho<$0.004 deg$^{-2}$; by comparing their source fluxes with those in the NVSS catalogues the characteristic timescale is $\sim$ 15 years (labelled `Croft2010$^{T}$' in Figure 4). 
The most stringent limit placed on the snapshot rate of sources is set by \cite{Gal-Yam} to be $\rho<$ 1.5$\times10^{-3}$ deg$^{-2}$ for flux densities $>$6 mJy (labelled `GY2006$^{T}$' in Figure 4). Note, the FIRST survey has improved angular resolution (5$^{\prime\prime}$) when compared with NVSS (45$^{\prime\prime}$), therefore correct source association effects transient identification.  We do not state a characteristic time scale for the FIRST-NVSS comparison as both individual surveys took a number of years; specific timescales can only be considered on a source by source basis.

\cite{Keith} set a limit on the snapshot rate of transient sources (calculated from detections) at 0.848 GHz to be $\rho<1.3\times10^{-2}$ deg$^{-2}$ above 14 mJy at a variety of timescales (labelled `Ban2011$^{T}$' in Figure 4). For variable radio sources a rate of $\rho<0.268$ deg$^{-2}$ is expected between a flux density 14 to 100 mJy (labelled `Ban2011$^{V}$' in Figure 4).

The upper limit derived from this study is consistent with the detections reported by \cite{Bower_2007} -- we might have expected possibly one detection at our flux density thresholds, assuming that the transient population sampled is isotropically distributed. \cite{Bower_2007} did note that an overdensity of galaxies was found within their field. The rate derived from this work is also broadly consistent with that of \cite{Bower2010}, \cite{Keith} and \cite{BowerCAL}. If the \cite{Bower_2007} and \cite{Keith} detections are of a similar nature, then some of the surveys that report upper limits sit very close to the `real' Log N - Log S. This is the best benchmark to date to predict the parameter space a given survey should probe to find transients. However, measurements such as frequency dependence and characteristic timescale of transient behaviour still need to be constrained. 

\section{Predictions for LOFAR}

Commissioning observations are currently underway with LOFAR that are probing the parameter space described in this paper. 
We include in Figure 4 a currently theoretical upper limit of $\rho<$ 0.012 deg$^{-2}$ based on zero detections from 10 epochs of 12 hour observations, each of 25 deg$^{2}$ fields at 150 MHz (labelled LOFAR$^{C}$). Early observations with LOFAR around August - September 2010 yield a typical RMS of 15 mJy (75 mJy for a 5$\sigma$ detection which is plotted) with a bandwidth of 50 MHz spread over 256 sub-bands (16 channels per subband). However as more baselines have come on-line, and next-generation data reduction strategies have been implemented, improvements upon this value have been made.  
A more realistic final theoretical noise of 0.36 mJy based on 18 core and 18 remote stations is indicated in Figure 4 (labelled LOFAR$^{F}$). We also include a theoretical prediction based on 10 epochs of 419 deg$^{2}$ fields at 30 MHz (labelled `LOFAR$^{F}$ 30 MHz' ). This sets an upper limit on the snapshot rate of transients to $\rho<$ 0.0018 deg$^{-2}$ above a detection threshold 30 mJy (6 mJy RMS noise). 

Recent work by \cite{Lazio2011} using the Long Wavelength Demonstrator Array (LWDA) -- a 16 dipole phased array with all-sky imaging capabilities -- have performed an all-sky blind transient search. A total of 106 hours of data was searched for radio transients at 73.8 MHz - the largest survey yet (in imaging mode) at low frequencies. With no detections of radio transients outside of the solar system above a flux density 500 Jy, an upper limit of 10$^{-2}$ yr$^{-1}$ deg$^{-2}$ is placed on the rate of events. With a typical integration time of 5 minutes, this converts to $\rho<$ 9.5$\times10^{-8}$ deg$^{-2}$; we plot this limit in Figure 4 (labelled `Laz2010') assuming a 5$\sigma$ (2500 Jy) detection is needed.
The \cite{Lazio2011} results tell us that extremely bright radio transients with characteristic timescale $\sim$ 5 minutes are very rare. 
Observations with LOFAR at 30 MHz (see Figure 4) would be complementary to the \cite{Lazio2011} survey. Approximately 20 tiled pointings could offer the same solid angle coverage as the LWDA i.e. the whole sky, with increased sensitivity. Pushing into this parameter space on a logarithmically spaced range of timescales is a goal for LOFAR, as well as an all-sky monitoring functionality to catch the brightest and rarest exotica. 

We can see from Figure 4 that if LOFAR observations were separated $\sim$ weekly, we would be able -- via sampling similar parameter space to the \cite{Bower_2007} detections -- to test the differences between a GHz and MHz population of radio transients.
If the emission mechanism for the GHz population is predominantly via the synchrotron process, then many sources will be initially optically thick within the LOFAR band. The rise time for a distant, luminous, population of radio transients -- such as GRB afterglows -- could be months or years; with lower peak fluxes. Therefore a steep spectrum population of coherent emitters might dominate detections in the LOFAR band. This coherent population will not be limited by the brightness temperature limit of the synchrotron sources and they could also have more erratic cadences -- i.e. switch on and off -- which should in turn affect the snapshot rate of events. 

Over ten observations of the same field, with approximately a weekly cadence have already been obtained with LOFAR at 150 MHz. The data reduction is in progress and a concise transient search will follow soon. We therefore hope to test the hypothesis above shortly and push further using future experiments with LOFAR.    

\section{Conclusion}

In this paper we have presented the results of an archival VLA study. We have calibrated, imaged and searched 5037 images of the calibrator fields totalling 435 hours of observing time. We have presented the methodology for reducing VLA data in a pipeline procedure. We have also explored some of the false detections that can be produced from pipelined image reduction. It should be noted that in general for future surveys our transient detection algorithms should be capable of recognising (and flagging) common errors associated with interferometry imaging, hence reducing the number of false detections produced. For example, quality control measurements should be included in the pipeline that assess and extract measurements from the observation to remove images from further transient searching. 

Even in a very simplistic implementation, these could include, measuring the flux of the calibrator source(s) and removing images where the flux had deviated away from the correct flux, or ignoring any image where more than $>$ 100 sources (or any number more than expected) are extracted. In this survey, a number of candidate transient sources were found to lie on the dirty beam. Source extraction could be performed on the dirty beam and checked for associations in the CLEANed image. Both false and real transient sources can be expected to lie on the dirty beam, however, this information could be used to lower the significance of a given detection in further analysis. 

If false detections find their way into the database, they should be systematically removed to avoid chance source associations in future observations. More complex, and computationally intensive interrogations of the database should also be performed to find lower significance detections. Greater consideration should also be given to the automated flagging algorithms. For this survey flagged data was very minimal. For LOFAR, however, large amounts of data may be removed, due to the nature of low frequency RFI. Examining the flagged visibilities, or even imaging them, could yield transient detections. 

This survey did not detect any radio transients and we have placed a constraint on the snapshot rate of radio transients. We have compared this constraint with results from other surveys, and although we did not detect any transients it is clear that large volumes of parameter space still remain unexplored. As new surveys push into the sub-mJy regime with various cadences and at different frequencies, definitive transient source populations should become apparent. Therefore with the next generation of radio telescopes such as LOFAR becoming available soon, the problem of inadequate sampling of rare transient phenomena will be alleviated. Due also to the triggering of multi-wavelength followup the classification and interpretation of these events will come of age. 

It seems clear that a large population of bright ($>$ mJy), GHz, frequent events does not exist. 
Therefore the sub-mJy GHz regime is clearly an important part of parameter space to probe for radio transients. The high dynamic range, and wide-field capabilities, of GHz instruments such as APERTIF (APERture Tiles In Focus, wide field upgrade to the Westerbork Synthesis Radio Telescope; \citealt{Apertif}) and ASKAP make them attractive, potentially high yield, discovery instruments.  
In summary, a large population of faint (probably distant) transients, as well as very rare bursts remains a strong possibility. 
Going both deeper and wider with the next generation of radio facilities will allow us to test these possibilities. 

\section*{acknowledgements}
The National Radio Astronomy Observatory is a facility of the National Science Foundation operated under cooperative agreement by Associated Universities, Inc. The authors would like to thank the NRAO archive staff for making this work possible.
The authors would like to specifically thank John Benson at NRAO archive for delivering vast quantities of archival data. 
MB would like to thank Bryan Gaensler, Tara Murphy and Keith Bannister from the University of Sydney for their useful comments and discussions.

\appendix
\bsp

\label{lastpage}


\begin{thebibliography}{99}

\bibitem[Bannister et al.(2011)]{Keith} Bannister, K.~W., 
Murphy, T., Gaensler, B.~M., Hunstead, R.~W., Chatterjee, S. 2011, MNRAS, 31 

\bibitem[Bastian et al.(1988)]{Bast_Flare} Bastian, T.~S., Dulk, 
G.~A., \& Slee, O.~B.\ 1988, AJ, 95, 794

\bibitem[Becker et al.(1995)]{Becker_FIRST} Becker, R.~H., White, 
R.~L., Helfand, D.~J. 1995, ApJ, 450, 559 

\bibitem[Becker et al.(2010)]{Becker} Becker, R.~H., Helfand, 
D.~J., White, R.~L., Proctor, D.~D. 2010, AJ, 140, 157 

\bibitem[Bell et al.(2010)]{Me} Bell, M.~E., et al.\ 2010, 
MNRAS, 1825 

\bibitem[Bhat et al.(2008)]{GP} Bhat, N.~D.~R., Tingay, 
S.~J., \& Knight, H.~S. 2008, ApJ, 676, 1200 

\bibitem[\protect\citeauthoryear{Boncz}{2003}]{Monet} P.A.~Boncz, \textsl{Monet: A Next-Generation DBMS Kernel For Query-Intensive Applications},
PhD Thesis, University of Amsterdam, 2002

\bibitem[\protect\citeauthoryear{Booth et al.}{2009}]{MeerKAT} 
Booth R.~S., de Blok W.~J.~G., Jonas J.~L., Fanaroff B., 2009, arXiv, 
arXiv:0910.2935 

\bibitem[\protect\citeauthoryear{Bower et al.}{2003}]{Bower_Orion} 
Bower G.~C., Plambeck R.~L., Bolatto A., McCrady N., Graham J.~R., de Pater 
I., Liu M.~C., Baganoff F.~K., 2003, ApJ, 598, 1140 

\bibitem[\protect\citeauthoryear{Bower et al.}{2005}]{Bower_GC_2005} 
Bower G.~C., Roberts D.~A., Yusef-Zadeh F., Backer D.~C., Cotton W.~D., 
Goss W.~M., Lang C.~C., Lithwick Y., 2005, ApJ, 633, 218 

\bibitem[\protect\citeauthoryear{Bower et al.}{2007}]{Bower_2007} 
Bower G.~C., Saul D., Bloom J.~S., Bolatto A., Filippenko A.~V., Foley 
R.~J., Perley D., 2007, ApJ, 666, 346

\bibitem[Bower et al.(2010)]{Bower2010} Bower, G.~C., et al.\ 
2010, ApJ, 725, 1792 

\bibitem[Bower 
\& Saul(2010)]{BowerCAL} Bower, G.~C., \& Saul, D.\ 2010, arXiv:1101.0121 

\bibitem[\protect\citeauthoryear{Briggs}{1995}]{Briggs} Briggs, B. S. \textsl{High Fidelity Deconvolution of Moderately Resolved Sources},
 PhD Thesis, The New Mexico Institute of Mining and Technology , 1995

\bibitem[Carilli et al.(2003)]{Carilli} Carilli, C.~L., Ivison, 
R.~J., Frail, D.~A. 2003, ApJ, 590, 192 

\bibitem[Cohen et al.(2007)]{VLSS} Cohen, A.~S., Lane, 
W.~M., Cotton, W.~D., Kassim, N.~E., Lazio, T.~J.~W., Perley, R.~A., 
Condon, J.~J.,  Erickson, W.~C.\ 2007, AJ, 134, 1245 

\bibitem[Condon et al.(1998)]{Condon_NVSS} Condon, J.~J., Cotton, 
W.~D., Greisen, E.~W., Yin, Q.~F., Perley, R.~A., Taylor, G.~B., 
 Broderick, J.~J. 1998, AJ, 115, 1693 

\bibitem[\protect\citeauthoryear{Cordes, Lazio, 
\& McLaughlin}{2004}]{Cordes_2004} Cordes J.~M., Lazio T.~J.~W., McLaughlin M.~A., 2004, NewAR, 48, 1459

\bibitem[Croft et al.(2010)]{Croft} Croft, S., et al.\ 2010, 
arXiv:1006.2003 

\bibitem[\protect\citeauthoryear{Davies et al.}{1976}]{Davies} 
Davies R.~D., Walsh D., Browne I.~W.~A., Edwards M.~R., Noble R.~G., 1976, 
Nature, 261, 476 

\bibitem[de Vries et al.(2004)]{devries} de Vries, W.~H., 
Becker, R.~H., White, R.~L., \& Helfand, D.~J.\ 2004, AJ, 127, 2565 

\bibitem[\protect\citeauthoryear{Eck, Cowan, 
\& Branch}{2002}]{eck_SN} Eck C.~R., Cowan J.~J., Branch D., 2002, ApJ, 573, 306 

\bibitem[Kuniyoshi et al.(2007)]{Kuniyoshi_2007} Kuniyoshi, M., et 
al. 2007, PASP, 119, 122 

\bibitem[\protect\citeauthoryear{Fender et al.}{2008}]{Fender_2008} 
Fender R., Wijers R., Stappers B., LOFAR Transients Key Science Project t., 
2008, arXiv, arXiv:0805.4349 

\bibitem[\protect\citeauthoryear{Frail et al.}{1997}]{Frail_GRB} 
Frail D.~A., Kulkarni S.~R., Nicastro L., Feroci M., Taylor G.~B., 1997, 
Nature, 389, 261 

\bibitem[Frail et al.(2003)]{Frail_CAT} Frail, D.~A., Kulkarni, 
S.~R., Berger, E., Wieringa, M.~H. 2003, AJ, 125, 2299 

\bibitem[\protect\citeauthoryear{Gaensler et 
al.}{2005}]{gaensler_magnetar} Gaensler B.~M., et al., 2005, Nature, 434, 
1104 

\bibitem[Gal-Yam et al.(2006)]{Gal-Yam} Gal-Yam, A., et al.
2006, ApJ, 639, 331 

\bibitem[Greisen(2003)]{AIPS} Greisen, E.~W. 2003, 
Astrophysics and Space Science Library, 285, 109

\bibitem[\protect\citeauthoryear{Hessels et 
al.}{2009}]{Hessels_LOFAR} Hessels J.~W.~T., Stappers B.~W., van 
Leeuwen J., Transients Key Science Project L., 2009, arXiv, arXiv:0903.1447

\bibitem[\protect\citeauthoryear{Hodapp et al.}{2004}]{PanSTARRS} 
Hodapp K.~W., et al., 2004, AN, 325, 636 

\bibitem[H{\"o}gbom(1974)]{CLEAN} H{\"o}gbom, J.~A.\ 1974, aaps, 15, 417 

\bibitem[Hopkins et al.(2002)]{FDR} Hopkins, A.~M., Miller, 
C.~J., Connolly, A.~J., Genovese, C., Nichol, R.~C., 
\& Wasserman, L.\ 2002, AJ, 123, 1086 

\bibitem[\protect\citeauthoryear{Hyman et al.}{2002}]{Hyman_2002_GC} 
Hyman S.~D., Lazio T.~J.~W., Kassim N.~E., Bartleson A.~L., 2002, AJ, 123, 
1497 

\bibitem[\protect\citeauthoryear{Hyman et al.}{2009}]{hyman_recent_galactic_2009} 
Hyman S.~D., Wijnands R., Lazio T.~J.~W., Pal S., Starling R., Kassim 
N.~E., Ray P.~S., 2009, ApJ, 696, 280

\bibitem[\protect\citeauthoryear{Johnston et 
al.}{2008}]{ASKAP} Johnston S., et al., 2008, ExA, 22, 151 

\bibitem[Jones et al.(2010)]{Sadie} Jones, S., McHardy, I., 
Moss, D., Seymour, N., Breedt, E., Uttley, P., Kording, E., 
\& Tudose, V.\ 2010, arXiv:1011.6633 

\bibitem[Kettenis et al.(2006)]{Kettenis} Kettenis, M., van 
Langevelde, H.~J., Reynolds, C., 
 Cotton, B. 2006, Astronomical Data Analysis Software and Systems XV, 351, 497 

\bibitem[\protect\citeauthoryear{Kida et al.}{2008}]{kida} 
Kida S., et al., 2008, NewA, 13, 519 

\bibitem[Lazio et al.(2010)]{Lazio2011} Lazio, T.~J.~W., et al.\ 
2010, AJ, 140, 1995 

\bibitem[\protect\citeauthoryear{Lenc et al.}{2008}]{Lenc_Trans} 
Lenc E., Garrett M.~A., Wucknitz O., Anderson J.~M., Tingay S.~J., 2008, 
ApJ, 673, 78 

\bibitem[Levinson et al.(2002)]{2002ApJ...576..923L} Levinson, A., Ofek, 
E.~O., Waxman, E., \& Gal-Yam, A.\ 2002, ApJ, 576, 923 

\bibitem[\protect\citeauthoryear{Lonsdale et 
al.}{2009}]{MWA} Lonsdale C.~J., et al., 2009, arXiv, 
arXiv:0903.1828 

\bibitem[Matsumura et al.(2009)]{Matsumura_2009} Matsumura, N., et 
al. 2009, \aj, 138, 787 

\bibitem[\protect\citeauthoryear{McLaughlin et 
al.}{2006}]{RRAT_1} McLaughlin M.~A., et al., 2006, Nature, 
439, 817 

\bibitem[Niinuma et al.(2007)]{Niinuma_2007} Niinuma, K., et al. 
2007, \apjl, 657, L37 

\bibitem[Niinuma et al.(2009)]{Niinuma_2009} Niinuma, K., et al.
2009, \apj, 704, 652 

\bibitem[\protect\citeauthoryear{Ofek et al.}{2009}]{Ofek} 
Ofek E.~O., Breslauer B., Gal-Yam A., Frail D., Kasliwal M.~M., Kulkarni 
S.~R., Waxman E., 2009, arXiv, arXiv:0910.3676 

\bibitem[\protect\citeauthoryear{Rau et al.}{2009}]{PTF} 
Rau A., et al., 2009, arXiv, arXiv:0906.5355 

\bibitem[Rengelink et 
al.(1997)]{WENSS} Rengelink, R.~B., Tang, Y., de Bruyn, A.~G., Miley, G.~K., Bremer, M.~N., Roettgering, H.~J.~A., \& Bremer, M.~A.~R.\ 1997, A\&A, 124, 259

\bibitem[\protect\citeauthoryear{Scheers}{2010}]{Bart} Scheers, B. \textsl{Transient and Variable Radio Sources in the LOFAR Sky. An Architecture for a Detection Framework.}, PhD Thesis, University of Amsterdam, 2010

\bibitem[\protect\citeauthoryear{Spreeuw}{2010}]{Hanno} Spreeuw, J. N. \textsl{Search and detection of low frequency radio transients},
 PhD Thesis, University of Amsterdam, 2010

\bibitem[\protect\citeauthoryear{Swinbank}{2007}]{Swinbank} 
Swinbank J., 2007, in Tzioumis T., ed., Proc. Bursts, Pulses and Flickering:
Wide-Field Monitoring of the Dynamic Radio Sky. p. 44

\bibitem[\protect\citeauthoryear{van den Oord 
\& de Bruyn}{1994}]{Flare_oort} van den Oord G.~H.~J., de Bruyn A.~G., 1994, A\&A, 286, 181

\bibitem[Verheijen et al.(2008)]{Apertif} Verheijen, M.~A.~W., 
Oosterloo, T.~A., van Cappellen, W.~A., Bakker, L., Ivashina, M.~V., 
\& van der Hulst, J.~M.\ 2008, The Evolution of Galaxies Through the Neutral Hydrogen Window, 1035, 265 

\bibitem[\protect\citeauthoryear{Welch et al.}{2009}]{ATA} 
Welch J., et al., 2009, arXiv, arXiv:0904.0762 

\bibitem[White et al.(1989)]{Flare_stars} White, S.~M., Jackson, 
P.~D., \& Kundu, M.~R.\ 1989, ApJS, 71, 895 

\bibitem[White et al.(1997)]{White_FIRST} White, R.~L., Becker, 
R.~H., Helfand, D.~J., Gregg, M.~D.\ 1997, \apj, 475, 479 

\bibitem[\protect\citeauthoryear{Zhao et al.}{1992}]{Zhao} 
Zhao J.-H., et al., 1992, Sci, 255, 1538 

\end{thebibliography}
\end{document}